\documentclass[showpacs,preprintnumbers,twocolumn,showkeys,superscriptaddress]{revtex4}
\usepackage{amssymb}
\usepackage[centertags]{amsmath}
\usepackage[paperwidth=210mm,paperheight=297mm,centering,hmargin=1.8cm,vmargin=2.0cm]{geometry}
\usepackage{txfonts}
\usepackage{epsfig}
\usepackage{bm}
\usepackage{graphicx,graphics}

\begin{document}

\title{Baryon-antibaryon production asymmetry in relativistic heavy ion collisions }
\author{Jun Song }
\affiliation{Department of Physics, Jining University, Shandong
273155, People's Republic of China}
\author{Feng-lan Shao}
\email{shaofl@mail.sdu.edu.cn} \affiliation{Department of Physics, Qufu Normal
University, Shandong 273165, People's Republic of China}

\begin{abstract}
Experimental data at RHIC and SPS energies suggest a systematic correlation between yield ratio $\bar{p}/p$ and $K^{-}/K^{+}$, which reveal some universality of hadron production in relativistic heavy ion collisions. We propose an explanation based on the quark combination mechanism in which the production asymmetry between baryons and antibaryons is focused on especially.  We start from the basic ideas of quark combination to carry out a general analysis of the properties of the global production of baryons and mesons and obtain the yields of baryons, antibaryons and mesons as functions of the number of constituent quarks and antiquarks just before hadronization. In particular, we study effects of quark-antiquark asymmetry on the ratio of (anti-)baryons to mesons and that of antibaryons to baryons. We use these formula to well explain the data of $\bar{p}/p$ and $K^{-}/K^{+}$ at different of collision energies and rapidities and suggest further measurements at moderate RHIC energies in the mid-forward rapidity regions to make the decisive test of this universality.
\end{abstract}

\maketitle

\section{Introduction}
Heavy-ion collisions at relativistic energies produce a long lifetime bulk quark matter with extremely high temperature and energy density \cite{Kolb0305084nuth}.
The system has a nonzero baryon number density due to the ever present nuclear stopping \cite{bearden:2004stop}.
This leads to the asymmetry between the production of hadrons and that of antihadrons, measured mainly by the ratios of yields of antihadrons to hadrons \cite{bearden:2003}.
Production of baryons and antibaryons, compared with that of mesons, is strongly influenced by the baryon number density of the system due to the baryon number conservation\cite{bearden:2003}.

Ratios of yields of antiprotons to protons and that of $K^{-}$ to $K^{+}$ are two of the most significant observables measuring the hadron-antihadron asymmetry in heavy ion collisions.
Final state proton ratio $\bar{p}/p$ carries the information of the production of baryons and antibaryons.
Kaon ratio $K^{-}/K^{+}$ roughly cancels the effect of strangeness production and reflects the asymmetry between charged mesons and their antiparticles.
Experimental data of this pair of baryon and meson ratios are available from the Relativistic Heavy Ion Collider (RHIC) \cite{bearden:2001,bearden:2003,arsene:2010,abel09pik,kuma11bes}, from the relatively low energy collisions at the Super Proton Synchrotron (SPS) and Alternating Gradient Synchrotron (AGS) \cite{ahle:1999,na49:2002,IGbearden:2002} and more recently from the very high energy reactions at the Large Hadron Collider (LHC) \cite{RPregh11PbPbLhc}.
The data seem to suggest a universal correlation between the $\bar{p}/p$ ratio and $K^{-}/K^{+}$ ratio at these collisional energies \cite{arsene:2010}.
Popular event generators A Multiphase Transport Model (AMPT) default version \cite{LinAMPT05} and Ultra-Relativistic
Quantum Molecular Dynamics (UrQMD) \cite{urqmd98review}, both dealing hadronization with string fragmentation, cannot explain this phenomenon \cite{arsene:2010}.
We note that the results of these two generators are similar to that of PYTHIA for $pp$ collisions.
This indicates that hadronic rescattering effects in heavy ion collisions have little relevance to the observed correlation which should be closely related to the hadronization of bulk quark matter.

In this paper, we use quark combination mechanism (QCM) to explain the observed correlation between $\bar{p}/p$ and $K^{-}/K^{+}$ ratio. Quark combination attributes the ratios of antihadrons to hadrons to the asymmetry between the number of constituent quarks and that of antiquarks before hadronization. ${K^{-}}/{K^{+}}$ ratio is mainly determined by $ {N_{\bar{q}}}/{N_{q}}$ because of strangeness neutrality $N_s=N_{\bar{s}}$ while $ {{\bar{p}}}/{p}$ ratio depends on the high power of $ {N_{\bar{q}}}/{N_{q}}$, e.g. ${{\bar{p}}}/{p} \sim \left({N_{\bar{q}}}/{N_{q}}\right)^3$ in a naive quark counting. Here, $q$ and $\bar{q}$ denote light quark and antiquark, respectively. In this view, the final state $\bar{p}/p$ ratio as the function of $K^{-}/K^{+}$ ratio has an intuitive physical meaning --- it reveals the law of the production asymmetry between baryons and antibaryons with respect to the baryon number density of the system characterized by the quark-antiquark asymmetry.
Therefore, we will emphasize on the bulk production of baryons and antibaryons in QCM.

The paper is organized as follows. In Sec II, we will start from the basic ideas of the combination mechanism to carry out a general analysis of the yield of baryons and that of mesons. We study the dependence of yields of directly produced baryons and mesons at hadronization on the number of constituent quarks and antiquarks. In particular, we are interested in the influence of quark-antiquark asymmetry on the yield ratio of (anti-)baryons to mesons and that of antibaryons to baryons. We will make the study as independent of the particular models as possible but present the assumptions and/or inputs explicitly whenever necessary.
In Sec III, we make the comparison with the experimental data of the $\bar{p}/p$ ratio with respect to $K^{-}/K^{+}$ ratio in heavy ion collisions at energies ranging from SPS to LHC. Sec IV summaries our work.

\section{General analysis of hadronic yields in QCM}

Considering a bulk quark system consisting of $N_q$  constituent quarks and $N_{\bar{q}}$ antiquarks, after hadronization the system changes the basic degrees of the freedom to become the hadronic matter and produces in average
\begin{description}
  \centering
  \item  $B(N_q,N_{\bar{q}})$ baryons,
  \item   \ \ \ \ $\bar{B}(N_q,N_{\bar{q}})$ antibaryons,
   \item   $M(N_q,N_{\bar{q}})$ mesons
\end{description}
which are functions of $N_q$ and  $N_{\bar{q}}$. Here we temporarily leave the hierarchy properties of quarks such as flavor and mass aside to focus on the global production of baryons and mesons. $M$ and $B(\bar{B})$ refer to the number of all mesons and (anti-)baryons, respectively.

An important property of yield functions of hadrons comes from the charge conjugation transformation of quark to antiquark. Under this transformation, baryons become antibaryons, antibaryons become baryons and mesons still keep mesons. Hadron yields change as $B \leftrightharpoons \bar{B}$  and $M \leftrightharpoons M$ under $q \leftrightharpoons \bar{q}$ exchange, and thus we have
\begin{equation}
\begin{split}
B(N_q,N_{\bar{q}}) =\bar{B}(N_{\bar{q}},N_q),  \\
M(N_q,N_{\bar{q}}) = M(N_{\bar{q}},N_q).
\end{split}
\label{exchange}
\end{equation}
%This exchange symmetry greatly constrains the functional forms of yields of baryons and mesons.

Quark combination mechanism describes the production of hadrons by the combination of constituent quarks and antiquarks, that is, a quark and an antiquark merge into a meson and three quarks into a baryon.
After system hadronization, there are no free quarks and antiquarks left and all of them are combined into hadrons.
We use the following constituent quark number conservation to denote the finish of hadronization process,
\begin{equation}
\begin{split}
   M(N_q,N_{\bar{q}})+3B(N_q,N_{\bar{q}})=N_q ,\\
   M(N_q,N_{\bar{q}})+3\bar{B}(N_q,N_{\bar{q}})=N_{\bar{q}}.
\end{split}
\label{unitarity}
\end{equation}
This unitarity is often adopted in studying hadronic yields in literatures \cite{alcor1995,Ycb06,Lzw11}. The related issues of entropy and energy conservation for such an effective $2\rightarrow 1$ or $3\rightarrow 1$ process have been extensively discussed and partially addressed by considering volume expansion and temperature decrease during system hadronization, etc \cite{Greco04,muller05,hwa04,fries08review,biro2007entropy,SJentrop10}.
Here we neglect the production of multiquark states and exotic hadrons. Since we are interested in hadronic yields which are low $p_T$ dominated, we also neglect the hadronization of jet quarks and the interaction of jet quarks and thermal quarks in the matter because of their negligible contributions to hadronic yield.

In addition, as the system entirely consists of quarks or of antiquarks, quark combination, if still works, should give
\begin{equation}
\begin{cases}
  \bar{B}=0, \ \  B=\frac{N_q}{3}, M=0         & \text{if } N_{\bar{q}}=0 \\
  \bar{B}=\frac{N_{\bar{q}}}{3},   B=0, \  \ M=0 &\text{if } N_{q}=0
\end{cases}
\label{boundary}
\end{equation}
These are boundary conditions of yield function of baryons and mesons in quark combination.

Obviously, the yield of baryons/antibaryons under boundary conditions Eq.(\ref{boundary}) is the linear function of that of quarks/antiquarks. In general case of both mesons and baryons being produced, supposing that we synchronously increase (or decrease) the number of quarks and that of antiquarks to vary the size of quark system at hadronization without changing the intensive properties of the system such as temperature and baryon number density, intuitively, hadronic yields should vary with the corresponding magnitude, i.e. linearity holds also. This linearity is supported by the RHIC data which show that the multiplicity of charged particles in central and semi-central nucleus-nucleus collisions is of nearly linear proportion to the number of participant nucleons $N_{part}$ \cite{bback06nch} while the ${\bar{p}/\pi^{-}}$ yield ratio, even at LHC energies \cite{RPregh11PbPbLhc}, is almost independent of $N_{part}$ \cite{abel09pik,adler04pkpi}. Calculations of the quark combination model developed by Shandong Group (SDQCM) \cite{Shao2005prc} also support such a property.
Letting $\lambda$ is a not very small auxiliary parameter, we thus have
\begin{equation}
\begin{split}
   M(\lambda N_q,\lambda N_{\bar{q}})=\lambda   M(N_q,N_{\bar{q}}), \\
   B(\lambda N_q,\lambda N_{\bar{q}})=\lambda  B(N_q,N_{\bar{q}}), \\
   \bar{B}(\lambda N_q,\lambda N_{\bar{q}})=\lambda  \bar{B}(N_q,N_{\bar{q}}),
\end{split}
\label{linearity}
\end{equation}
in quark combination hadronization.
Note that here we focus on the bulk matter with abundant particles and do not consider the situation of small system with few quarks where extra restraints in production of hadrons, in particular of baryons and anti-baryons, maybe come out.

These properties Eqs. (\ref{exchange}-\ref{linearity}) pose important constraints on the behaviors of yields of baryons and mesons. First, we rewrite the arguments of yield function, i.e., $N_{q}$ and $N_{\bar{q}}$, in terms of $x$ and $z$ by
\begin{equation}
    N_q + N_{\bar{q}} =x,  \hspace{1.5cm}
    \frac{N_q - N_{\bar{q}}}{N_q + N_{\bar{q}}} =z.
\label{vxy}
\end{equation}
Variable $x$ is the total number of quarks and antiquarks, which characterizes the bulk property of the system related to the system size or energy. Variable $z$ depicts the asymmetry between quarks and antiquarks in the system ($|z|\le 1$), which is a measure of baryon number density of the system.
Under $q \leftrightharpoons \bar{q}$ permutation $x \leftrightharpoons x$  and $z \leftrightharpoons -z$ and yields of baryons and mesons can be written as the function of $x$ and $z$
\begin{eqnarray}
     B(N_q,N_{\bar{q}}) &=& B(x,z),  \label{bxz} \\
     \bar{B}(N_q,N_{\bar{q}})&=&\bar{B}(x,z)=B(x,-z), \label{bbxz} \\
     M(N_q,N_{\bar{q}})&=&M(x,z)=M(x,-z), \label{mxz}
\end{eqnarray}
where we have used the symmetry properties Eq. (\ref{exchange}). Correspondingly the linearity property Eq. (\ref{linearity}) can be expressed with variables $x,z$ as
\begin{equation}
\begin{split}
     M(\lambda x,z)=\lambda   M(x,z), \\
  B(\lambda x,z)=\lambda   B(x,z), \label{lambxz}\\
   \bar{B}(\lambda x,z)=\lambda \bar{B}(x,z).
\end{split}
\end{equation}
Since the $B(x,z)$ is the linear function of variable $x$,
we immediately have
\begin{equation}
  B(x,z)= x \mathcal{B}(z),  \label{xbz}
\end{equation}
and correspondingly for antibaryons and mesons,
\begin{eqnarray}
 \bar{B}(x,z)&=&B(x,-z)=x\,\mathcal{B}(-z), \label{xbbz}\\
 M(x,z) &=&M(x,-z)=\frac{1}{2}x\Big( 1-3\mathcal{B}(z)-3\mathcal{B}(-z) \Big).
  \label{xmz}
\end{eqnarray}
Properties of unitarity and boundary conditions at hadronization exert constraints on $\mathcal{B}(z)$ as follows
\begin{equation}
    \mathcal{B}(1)=\frac{1}{3}, \hspace{0.6cm} \mathcal{B}(-1) =0, \hspace{0.6cm}
      \mathcal{B}(z)-\mathcal{B}(-z)=z/3 .  \label{bzconstraint}
\end{equation}

Using Eqs. (\ref{xbz}) and (\ref{bzconstraint}) we obtain the ratio of yield of anti-baryons to that of baryons.
\begin{equation}
  R(z)=\frac{\bar{B}(x,z)}{B(x,z)}=\frac{\mathcal{B}(-z)}{\mathcal{B}(z)}=1-\frac{z}{3\,\mathcal{B}(z)}. \label{rz}
\end{equation}
The ratio $R$ satisfies
\begin{equation}
\begin{split}
   R(0)=1 ,        \\
  R(1)=0 ,        \\
  R(-1)=\infty ,  \\
  R(z)R(-z)=1 .  \label{rzbound}
\end{split}
\end{equation}
We can see that the behavior of $R(z)$ is greatly constrained by properties Eqs.(\ref{exchange}-\ref{linearity}) which come from the global properties and/or symmetries.

The complete determination of $R(z)$ needs further consideration of hadronization dynamics. In QCM, baryon is formed by the combination of three (anti-)quarks in neighborhood in the phase space. The (anti-)baryon yield is determined by two ingredients: (1) the probability of finding three neighboring (anti-)quarks in phase space $P_{qqq}$ ($P_{\bar{q}\bar{q}\bar{q}}$). (2) the probability of these three (anti)quarks to form a (anti-)baryon, $P_{qqq\rightarrow B }$ ($P_{\bar{q}\bar{q}\bar{q} \rightarrow \bar{B} }$).
The contribution of (1) to $R(z)$, $P_{\bar{q}\bar{q}\bar{q}}/ P_{qqq}$, follows the form of $(N_{\bar{q}}/N_q)^a$ for the system consists of free quarks and antiquarks.
The combination probability $P_{qqq\rightarrow B }$ ($P_{\bar{q}\bar{q}\bar{q} \rightarrow \bar{B} }$) in (2) is determined by the local hadronization dynamics.
If the fast hadronization approximation is assumed as did by most of combination models, the $P_{qqq\rightarrow B }$ is evaluated by the overlap between the hadronic wave function and quark ones, which is undisturbed by surrounding environment. Then it contributes nothing to $R(z)$ and we can obtain $R(z)=(N_{\bar{q}}/N_q)^a$.
On the other hand, if we consider the possible effects of the surrounding environment (i.e. quarks and antiquarks) to $qqq \rightarrow B$ process, we may introduce a further contribution to $R(z)$ because the surrounding quarks and antiquark are asymmetrical. Supposed that there is a $\bar{q}$ in the neighbor of $qqq$ cluster, this $\bar{q}$ can have the chance to capture one $q$  to form a meson and left two quarks combining with other quarks/antiquarks. On the other hand, if the nearest neighbor of $qqq$ cluster is a quark, the baryon formation should be then dominated.
Therefore, the ingredient (2) can also contribute to $R(z)$ via the form of $N_{\bar{q}}/N_q$.
We note that such a consideration is essentially a kind of dynamical competition between the production of baryon and that of meson.
Finally, we  can expect the following form of $R(z)$
\begin{equation}
  R(z)=\Big( \frac{1-z}{1+z} \Big )^{a}
\label{Rzpar}
\end{equation}
by noticing the identity $N_{\bar{q}}/N_q=(1-z)/(1+z)$. It is the simplest functional form with one parameter $a$ which satisfies Eq. (\ref{rzbound}).
This form can be tested by an intuitive combination process in the following way.  First, we sort all quarks and antiquarks in the system into a one dimensional quark sequence in which the position of quarks can denote the degree of proximity to each other. Second, we combine those $\bar{q}q$ and $qqq$ in neighbors into mesons and baryons, respectively, until all quarks and antiquarks are exhausted.

Parameter $a$ bears important hadronization dynamics which is reflected by
\begin{equation}
  a= \frac{1}{3 R_{B/M}(0)} +1  \nonumber
\end{equation}
where $R_{B/M}(0)={B(x,0)}/{M(x,0)}$ is the ratio of the yield of baryons to that of mesons at vanishing quark antiquark asymmetry $z=0$. This ratio denotes the competition between the formation of baryon and that of meson when quarks are combined into a hadron.  The study of hadron yields in Ref.\cite{wangrq2012} shows that $R_{B/M}(0)$ of value about $1/12$ can well describe the data of $\bar{p}/\pi^{-}$ yield ratio in Pb+Pb collisions at $\sqrt{s_{NN}}= 2.76$ TeV after including effects of resonance decay. Then we obtain $a\approx5$.

\begin{figure}[!hbtp]
  \includegraphics[width=8cm]{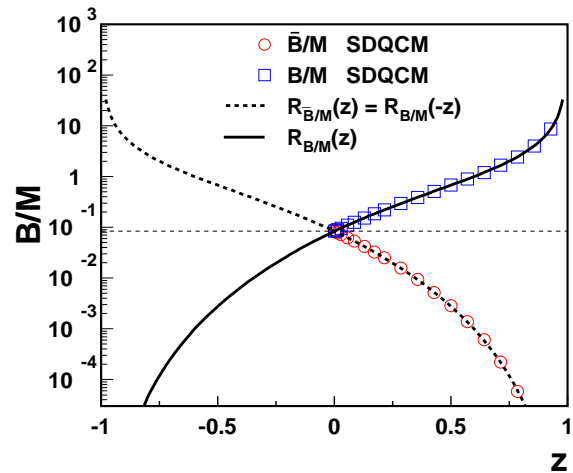}\\
  \caption{(Color online) The ratio of yield of (anti-)baryons to that of mesons as the function of $z$.}\label{bmratio}
\end{figure}

The functional forms of $B(x,z)$ and $\bar{B}(x,z)$ then read as
\begin{eqnarray}
  B(x,z)=x\,\mathcal{B}(z)=\frac{x\,z}{3}\frac{(1+z)^{a}}{(1+z)^{a}-(1-z)^{a}},\\
  \bar{B}(x,z)=x\,\mathcal{B}(-z)=\frac{x\,z}{3}\frac{(1-z)^{a}}{(1+z)^{a}-(1-z)^{a}}.
\end{eqnarray}
Figure \ref{bmratio} shows the results of $ R_{B/M}(z)$ and $ R_{\bar{B}/M}(z)= R_{B/M}(-z)$.  Symbols are numerical results of SDQCM \cite{Shao2005prc}.

\section{Comparison with experimental data }
In this section, we test our baryon formula by the available experimental data of $\bar{p}/p$ at RHIC and SPS. Since quark asymmetry $z$ is unobservable we should firstly relate it to an observation, ${N_{K^{-}}}/{N_{K^{+}}}$ yield ratio. Now we consider that (anti-)quarks have three different flavors, i.e. $u$, $d$ and $s$.   Their numbers are denoted by $N_{u,\bar{u}},N_{d,\bar{d}} \,\text{ and} \, N_{s,\bar{s}}$. Under isospin symmetry we can write $z$ as
\begin{equation}
  z=\frac{N_q-N_{\bar{q}}}{N_q+N_{\bar{q}}}=\frac{N_u - N_{\bar{u}}}{N_u + N_{\bar{u}}+ N_{{s}}}=\frac{1- \frac{N_{\bar{u}}}{N_u} }{1 + \frac{N_{\bar{u}}}{N_u} + \frac{N_{{s}}}{N_u} }.
\end{equation}
In the quark combination mechanism, ratios ${N_{\bar{u}}}/{N_u}$ and ${N_{{s}}}/{N_u}$ can be related to the corresponding ratio of different hadron species that contain these constituent quarks. For example, we have ${N_{K^{-}(\bar{u}s)}}/{N_{K^{+}(u\bar{s})}}\propto {N_{\bar{u}}}/{N_{u}}$ and ${N_{\phi(s\bar{s})}}/{N_{K^{+}(u\bar{s})}}\propto {N_{{s}}}/{N_{u}}$ in general because of the strangeness neutrality$N_{s}=N_{\bar{s}}$. Adopting our terminology of the description of quark combination in Ref.\cite{wangrq2012} for the proportion coefficients, these two ratios read as
\begin{equation}
  \frac{N_{K^{-}}}{N_{K^{+}}}=\frac{C_{K^{-}}}{C_{K^{+}}}\frac{N_{\bar{u}}}{N_{u}}, \hspace{1cm}
  \frac{N_{\phi}}{N_{K^{+}}}=\frac{C_{\phi}}{C_{K^{+}}}\frac{N_{{s}}}{N_{u}}.
\end{equation}
Here, $C_{K^{\pm}}$ and $C_{\phi}$ are combination factors denoting the branch ratio of a given constituent quark composition (e.g., $u\bar{s}$) to form a specific hadron species (e.g., $K^{+}$). $C_{K^{-}}=C_{K^{+}}$ due to both hadrons are in the same $J^P = 0^-$ pseudoscalar nonet. $C_{\phi}/C_{K^{+}}=R_{VP}$ where $R_{VP}$ represents the ratio of the $J^P = 1^-$ vector mesons to the $J^P = 0^-$ pseudoscalar mesons of the same flavor composition.

The influence of resonance decay on ${N_{K^{-}}}/{N_{K^{+}}}$ ratio is small because of almost same decay contributions to $K^{\pm}$, and we still have
\begin{eqnarray}
  \frac{N^{(final)}_{K^{-}}}{N^{(final)}_{K^{+}}}=\frac{N_{\bar{u}}}{N_{u}}.
\end{eqnarray}
For ${N_{\phi}}/{N_{K^{+}}}$ ratio, decay of resonances contributes significantly the denominator, and thus we have
\begin{equation}
  \frac{N^{(final)}_{\phi}}{N^{(final)}_{K^{+}}}\approx\frac{N_{\phi}}{N_{K^{+}}+N_{K^{*}}}
  =\frac{R_{VP}}{1+R_{V/P}}\frac{N_{{s}}}{N_u}.
\end{equation}
We further rewrite l.h.s of the equation as   $\frac{N^{(final)}_{\phi}}{N^{(final)}_{K^{-}}}\frac{N^{(final)}_{K^{-}}}{N^{(final)}_{K^{+}}}$ by noticing that the experimental data of ${N^{(final)}_{\phi}}/{N^{(final)}_{K^{-}}}$ is an almost collision-energy independent value, which is about 0.14 \cite{abelev:2009}. Then we can relate the argument $z$ to ${N^{(final)}_{K^{-}}}/{N^{(final)}_{K^{+}}}$ (denoted by ${K^{-}}/{K^{+}}$ for short) as following
\begin{equation}
  z=\frac{1-\frac{K^{-}}{K^{+}}}{1+\left(1+0.14\frac{1+R_{VP}}{R_{V/P}}\right)\frac{K^{-}}{K^{+}}}.
  \label{zvsKratio}
\end{equation}
The variance of $R_{V/P}$ exerts little influence on the transition between $z$ and ${K^{-}}/{K^{+}}$. Here we take $R_{V/P}=0.45$ based on the study of $K^{*}$ production in relativistic heavy ion collisions in Ref.\cite{zkkstar2012}.

By definition, $R(z)$ is the ratio of the yield of directly-produced anti-baryons to that of baryons in the system at hadronization. Since all (anti-)baryons will decay into (anti-)nucleons at last, $R(z)$ equals the ratio of final state anti-nucleons to that of nucleons and finally we obtain
\begin{equation}
R(z)= N^{final}_{\bar{p}}/N^{final}_{p}
\label{rzdata}
\end{equation}
under the isospin symmetry, which is denoted by $\bar{p}/p$ for short. Now, with Eqs. (\ref{Rzpar}) and (\ref{zvsKratio}-\ref{rzdata}) we can use the data of ${K^{-}}/{K^{+}}$ to predict the final state $\bar{p}/p$ ratio to test our formula of baryon production and underlying assumptions. Note that there is no difficulty of applying these global formula to the measured data at specific rapidities because the locality of hadronization also guarantees roughly the local flavor conservation especially for the case of quite high quark rapidity density in relativistic heavy ion collisions.

\begin{figure}[!tbhp]
  \includegraphics[width=8cm]{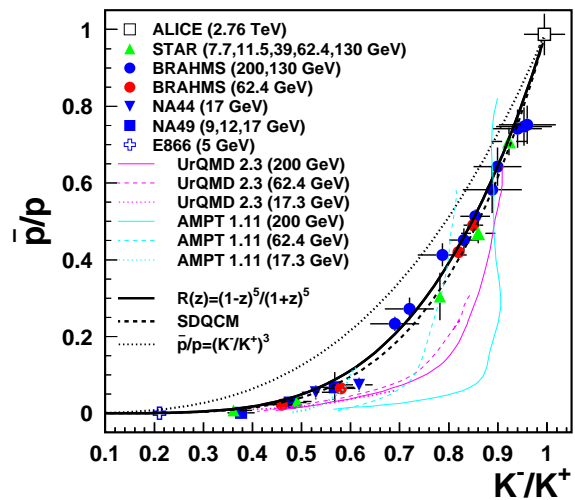}\\
  \caption{(Color online)  $\bar{p}/p$ yield ratio with respect to ${K^{-}}/{K^{+}}$ ratio in central nucleus-nucleus collisions. The symbols of RHIC energies are BRAHMS and STAR data from Refs. \cite{bearden:2001,bearden:2003,arsene:2010,abel09pik,kuma11bes} and lower energy data from Refs. \cite{ahle:1999,na49:2002,IGbearden:2002}. The data of LHC in central Pb+Pb collisions at $\sqrt{s_{NN}}= 2.76$ TeV are from Ref.\cite{RPregh11PbPbLhc}. Results of AMPT 1.11 and UrQMD 2.3 are taken from Ref. \cite{arsene:2010}. }
  \label{chemical_correlation}
\end{figure}

Figure \ref{chemical_correlation} shows the $\bar{p}/p$ ratio with respect to ${K^{-}}/{K^{+}}$ ratio in central nucleus-nucleus collisions at energies ranging from $\sqrt{s_{NN}}= 5 \textrm{ GeV}-2.76 \textrm{ TeV}$. The data points of RHIC energies \cite{bearden:2001,bearden:2003,arsene:2010,abel09pik,kuma11bes} are obtained in different rapidity slices, while the SPS and AGS points \cite{ahle:1999,na49:2002,IGbearden:2002} are obtained at midrapidity.
The data of LHC in central Pb+Pb collisions at $\sqrt{s_{NN}}= 2.76$ TeV \cite{RPregh11PbPbLhc} are also presented.
Comparing the low and high energy results, we observe a coincidence of the forward rapidity results of Au+Au 62.4 GeV and those at SPS energies, and also a coincidence of the midrapidity results of Au+Au 62.4 GeV and those of Au+Au 200 GeV at forward rapidities. This indicates the universality of hadron production in heavy ion collisions at these energies.
Results of event generators AMPT 1.11 with string fragmentation and UrQMD 2.3 in Fig. \ref{chemical_correlation} are taken from Ref. \cite{arsene:2010}.
They give an qualitative description for $\bar{p}/p$ variance with respect to ${K^{-}}/{K^{+}}$ ratio but a bad explanation for the universality of the correlation.
We note that results of these two generators are similar to that of PYTHIA for $pp$ collisions, which indicates that this systematic correlation should be relevant to the hadronization of bulk quark matter instead of hadronic rescattering effects in heavy ion collisions.

The solid line in Fig. \ref{chemical_correlation} is our result of ${\bar{p}}/{p}\approx R(z) $ by Eq. (\ref{Rzpar}) in terms of $z$ by Eq. (\ref{zvsKratio}). We observe a good agreement with experimental data.
The dashed line is the result of SDQCM which shows the influence of full treatment of resonance decay.
In our explanation, the exponent $a=5$ which determines the slope of $\bar{B}/B$ ratio is the inverse of the products of three factors $3\times2\times \mathcal{B}(0)$ by using the derivation of $R(z)$ at $z=0$, $\left. \frac{  \partial R(z) }{\partial z} \right|_{z=0}=-\frac{1}{3\,\mathcal{B}(0)}=-2a $. The 3 comes from the number of constituent quarks contained in baryon and the 2 from the symmetry property of baryon and antibaryon production. The $\mathcal{B}(0)$ is the proportion of the baryon yield in the hadronization of the bulk quark matter with zero baryon number density. They together predict the production properties of baryons and antibaryons at non-zero $z$.

We also show the argument of naive quark combination, i.e., ${{\bar{p}}}/{p}=\left({K^{-}}/{K^{+}}\right)^{3}$ by simple quark number counting ${K^{-}}/{K^{+}}\sim {N_{\bar{u}}}/{N_{u}}$ and ${{\bar{p}}}/{p} \sim \left({N_{\bar{u}}}/{N_{u}}\right)^3$. It is obviously deviated from (located above) the experimental data.
This is because such an inclusive point of view only considers the free combination of (anti-)quarks at the moment of (anti-)baryon production without the proper treatment of final fates of all other (anti-)quarks that are failing or are not participant in baryon formation.  The exclusive quark combination, as shown by our result, covers this unitarity of hadronization to yield the increased exponent $\alpha \approx 5$ for ${{\bar{p}}}/{p}=\left({K^{-}}/{K^{+}}\right)^{\alpha}$ parametrization, which is in good agreement with the data.
Description of baryon production in string fragmentation is significantly different from that in quark combination.
It gives a quite large exponent $\alpha \gtrsim 8 $  and resulting curves lie below the data in the intermediate $K^{-}/K^{+}$ region.

\section{summary}
We have studied the correlation between $\bar{p}/p$ and $K^{-}/K^{+}$ yield ratios at RHIC and SPS energies in the framework of the quark combination mechanism.
These two ratios can reflect the property of baryon production in relativistic heavy ion collisions.
Starting from the basic ideas of quark combination, we discussed the general properties of bulk hadron production in quark combination and obtained yields of baryons, antibaryons and mesons as functions of the numbers of constituent quarks and antiquarks before hadronization.
It is found that the observed correlation between $\bar{p}/p$ and $K^{-}/K^{+}$ ratios are well explained by our formula.
It is important to understand the low $p_T$ hadron production in relativistic heavy ion collisions.
We expect the Beam Energy Scan program of STAR Collaboration can complete the data set of $\bar{p}/p$ and $K^{-}/K^{+}$ ratio in the intermediate $z$ region to make a further test of this universality of hadron production in relativistic heavy ion collisions. The proper collision energy region for midrapidity measurements lie in  $\sqrt{s_{NN}}\approx 20-60$ GeV and for mid-forward rapidity measurements we can choose the rapidity region [1.0, 2,0] at $\sqrt{s_{NN}}=62.4$ GeV  and smaller rapidities at lower collision energies.

We note that these two ratios have been studied in statistical models \cite{becattini01,tawfik11} and can be explained well also. This kind of models use the chemical potential to explain ratios of antihadron to hadron and thereby gain insight into the equilibrium information of the hot and dense matter produced in relativistic heavy ion collisions. Our explanation of the correlation between $\bar{p}/p$ and ${K^{-}}/{K^{+}}$ ratio is based on a microscopic mechanism, which is different from those of statistical models.

\section*{ACKNOWLEDGMENTS}
The authors thank Q. B. Xie, Z. T. Liang, Q. Wang, G. Li and R. Q. Wang for helpful discussions.  The work is supported in part by the National Natural Science Foundation of China under grant 11175104 and 11247202, and by the Natural Science Foundation of Shandong Province, China under grant ZR2011AM006 and ZR2012AM001. J. Song acknowledges the hospitality of Professor Q. Wang and Doctors H. J. Xu and Y. K. Song in visiting USTC, China.

\end{document}